\documentclass[pre,preprintnumbers,twocolumn]{revtex4}
\usepackage{amssymb,amsmath,amsthm}
\usepackage[dvips]{graphicx}
\usepackage{verbatim}
\usepackage{epsfig}
\usepackage{color}
\usepackage{comment}

\begin{document}

%\title{Magneto-optic dynamics in a ferromagnetic nematic liquid crystal}
\title{Dynamic magneto-optic coupling in a ferromagnetic nematic liquid crystal}

\date{\today}

\author{ Tilen Potisk$^{1,2,*}$, Daniel Sven\v sek$^{1}$, Helmut R. Brand$^{2}$,
Harald Pleiner$^{3}$,\\
 Darja Lisjak$^{4}$, Natan Osterman$^{1,4}$, and Alenka Mertelj$^{4}$ }

\affiliation{$^{1}$Department of Physics, Faculty of Mathematics and Physics,
University of Ljubljana, SI-1000 Ljubljana, Slovenia\\
% $^{2}$Department of Physics, University of Bayreuth, 95440 Bayreuth, Germany \\
$^2$Theoretische Physik III, Universit\"at Bayreuth, 95440 Bayreuth, Germany \\
 $^{3}$Max Planck Institute for Polymer Research, 55021 Mainz, Germany\\
 $^{4}$J. Stefan Institute, SI-1000 Ljubljana, Slovenia \\
 $^{*}$e-mail: tilen.potisk@uni-bayreuth.de }

\begin{abstract}
\noindent 
Hydrodynamics of complex fluids with multiple order parameters is governed by a set of dynamic equations with many material constants, of which only some are easily measurable.
We present a unique example of a dynamic magneto-optic coupling in a ferromagnetic nematic liquid, in which long-range orientational order of liquid crystalline molecules is accompanied by long-range magnetic order of magnetic nanoplatelets.
We investigate the dynamics of the magneto-optic response 
experimentally and theoretically and find out that it is significantly affected by the dissipative dynamic cross-coupling between the nematic and magnetic order parameters.
The cross-coupling coefficient determined by fitting the experimental results with a macroscopic theory is of the same order of magnitude as the dissipative coefficient (rotational viscosity) that governs the reorientation of pure liquid crystals.

\end{abstract}
\maketitle
%\pacs{82.40.Bj, 05.70.Ln, 42.65.Sf, 47.20.Ky}

\noindent Fluids with an apolar nematic orientational ordering --
nematic liquid crystals (NLCs) -- are well known and understood, with properties
useful for different types of applications \cite{goodby}. As a
contrast, possible existence of fluids with a polar orientational
order, thus of a lower symmetry than NLCs, has been always intriguing.
An electrically polar nematic liquid was theoretically discussed \cite{born}
as early as 1910s, but such a phase has never been observed. Similarly,
vectorial magnetic ordering, i.e., ferromagnetism, is a phenomenon that
occurs in solids and has been for the longest time considered hardly
compatible with the liquid state.

Quite recently, however, ferromagnetic NLCs have been realized in suspensions of magnetic nanoplatelets in NLCs
\cite{alenkanature,shuaiNC,smalyukhPNAS} and their macroscopic static properties were
characterized in detail \cite{alenkasoftmatter}. These systems possess
two order parameters giving rise to two preferred directions -- the nematic director {\bf n} (denoting the average orientation of liquid crystalline molecules) and the spontaneous
magnetization $\bf M$ (describing the density of magnetic moments of the nanoplatelets) -- that are coupled statically as well as dynamically.
As a consequence, optical and magnetic responses are coupled in these
materials, which makes them particularly interesting in the multiferroic
context: optical properties can be manipulated with a weak external
magnetic field (a strong magneto-optic effect) and conversely, the
spontaneous magnetization can be reoriented by an external electric
field (the converse magnetoelectric effect). Note that subjecting the
liquid crystal to an external electric field is the usual means of
controlling the nematic director in optical applications.

The search for a ferromagnetic nematic phase started when Brochard
and de Gennes \cite{bropgdg} suggested and discussed a ferromagnetic
nematic phase combining the long range nematic orientational order
with long range ferromagnetic order in a fluid system. The synthesis
and experimental characterization of ferronematics and ferro\-cholesterics,
a combination of low-molecular-weight NLCs with
magnetic liquids leading to a superparamagnetic phase, started immediately
and continued thereafter \cite{rault,amer,kopcansky,ouskova,buluy,podoliak} (compare also Ref.~\cite{reznikov}
for a recent review). 
These studies were making use of ferrofluids
or magnetorheological fluids (colloidal suspensions of magnetic particles)
\cite{rosensweig}; their experimental properties \cite{rosensweig,odenbach2004}
have been studied extensively in modeling \cite{burylovII,mario2001,luecke2004,hess2006,sluckin2006,mario2008,klapp2015}
using predominantly macroscopic descriptions \cite{burylovII,mario2001,luecke2004,hess2006,sluckin2006,mario2008}.

On the theoretical side the macroscopic dynamics of ferronematics
was given first for a relaxed magnetization \cite{jarkova2} followed
by taking into account the magnetization as a dynamic degree of freedom
\cite{jarkova} as well as incorporating chirality effects leading
to ferrocholesterics \cite{fink2015}. In parallel a Landau description
including nematic as well as ferromagnetic order has been presented
\cite{hpmhd}.

In this Letter we describe experimentally and theoretically the dynamic
properties of ferromagnetic NLCs, focusing on the coupled evolution
of the magnetization and the director fields actuated by an external magnetic
field.
The dynamic coupling between $\bf M$ and $\bf n$ 
has been a complete blank to this day. It has been known
that it is allowed by symmetry and the rules of linear irreversible
thermodynamics, and was cast in a definite form theoretically 
\cite{jarkova} as a prediction.
Here we demonstrate that these coupling terms influence decisively
the dynamics. 
Quantitative agreement between the experimental results
and the model is reached and a dissipative cross-coupling coefficient
between the magnetization and the director is accurately evaluated.
It is shown that this cross-coupling is crucial to account
for the experimental results thus underscoring the importance of such
cross-coupling effects in this recent soft matter system. 
%The dynamic results shown in this Letter already for low magnetic fields demonstrate the potential for applications of ferromagnetic nematics in displays and magneto-optic devices as well as in the field of smart fluids.

\begin{figure}[htb]
\includegraphics[width=3.3in]{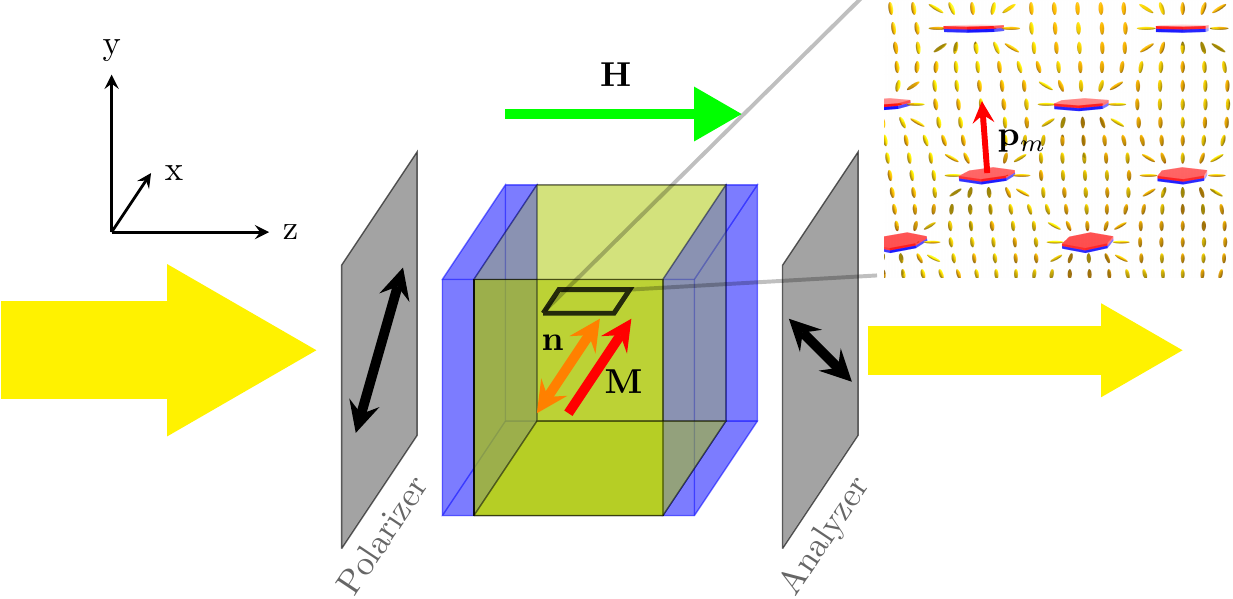} \caption{ (Color online) Sketch of the experimental set-up and definition of
coordinate axes. The thick yellow arrows indicate the direction of
the light passing through the polarizer and the analyzer. In the absence of an applied magnetic field
(${\bf H}$, $z$ direction), the equilibrium director (${\bf n}$)
and magnetization (${\bf M}$) fields are only slightly pretilted from the $x$ direction.
Inset: distortion of the NLC director (ellipsoids, schematic) prevents flocculation of the suspended nanoplatelets carrying a magnetic moment ${\bf p}_m$ parallel to $\bf n$ in equilibrium.}
\label{fig:scheme} 
\end{figure}

The suspension of magnetic nanoplatelets in the NLC pentylcyanobiphenyl (5CB, Nematel) was prepared as described
in Ref.~\cite{alenkasoftmatter}. The magnetic platelets with an average diameter of 70 nm and thickness
of 5 nm, made of barium hexaferrite doped with scandium, were covered by the surfactant dodecylbenzenesulphonic acid
(DBSA). The surfactant induced perpendicular anchoring of 5CB molecules on
the platelet surface leading to parallel orientation of the platelet
magnetic moments and the nematic director, Fig.~\ref{fig:scheme} (inset).
The volume concentration of the platelets in 5CB,
determined by measuring the saturated magnetization of the suspension, was
$\sim3\times10^{-4}$, which corresponds to the magnetization magnitude of
$M_{0}\sim50$\,A/m. 
The suspension was put in a liquid crystal cell with thickness $d\sim20\,\mu$m, inducing planar homogeneous
orientation of {\bf n} along the rubbing direction $x$, Fig.~\ref{fig:scheme}. 
During the filling process a magnetic field (not shown) of 8 mT was applied in the direction of the rubbing, so that a magnetic
monodomain sample was obtained. In the absence of an external magnetic field the spontaneous magnetization ${\bf M}$ was parallel to ${\bf n}$.

\begin{figure}[htb]
\includegraphics[width=3.3in]{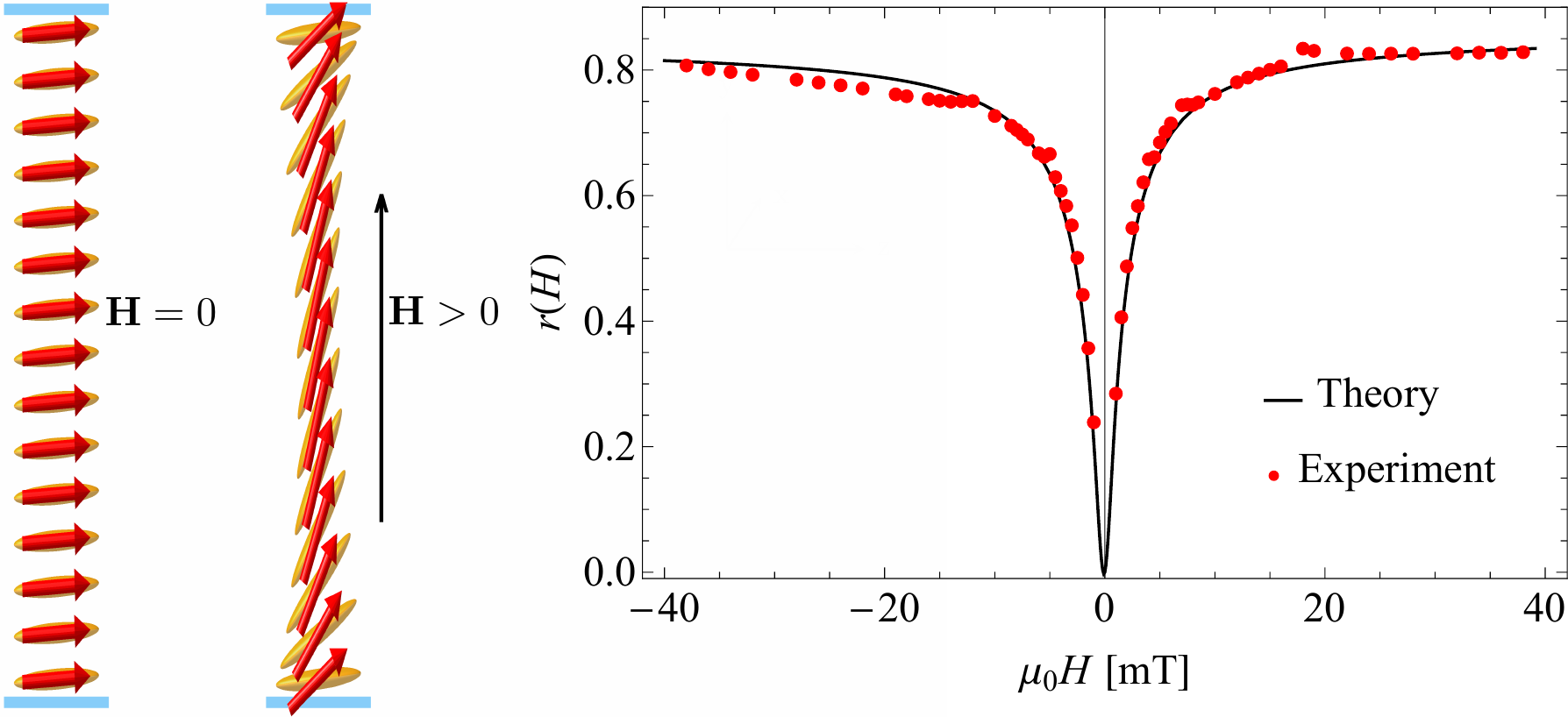} 
\caption{ (Color online) 
Left: response of the magnetization (red arrows) and the director (ellipsoids) to the external magnetic field $\bf H$ applied in the $z$ direction.
Right: equilibrium normalized phase difference $r(H)$ as a function of the magnetic field $\mu_{0}H$, fitted by the static model.}
\label{fig:statics} 
\end{figure}

The suspension exhibits a strong magneto-optic effect. For example, when a magnetic field $\bf H$ is applied perpendicularly to ${\bf M}$ ($z$ direction), it exerts a torque on the magnetic moments, i.e., on the platelets, and
causes their reorientation. Because the orientations of the platelets and the director are coupled through the anchoring of the NLC molecules on the platelet's surface, also ${\bf n}$ reorients, which is observed as an optic response. In Fig.~\ref{fig:statics} (left) the response of $\bf M$ and $\bf n$ is shown schematically; note the small angle between ${\bf M}$ and ${\bf n}$ in equilibrium.

The reorientation
of ${\bf n}$ is detected optically by measuring the phase difference $\phi$ between
transmitted extraordinary and ordinary light \cite{alenkasoftmatter}, Fig.~\ref{fig:scheme}. 
The normalized phase difference $r({H})=1-{\phi({H})/\phi_{0}}$, where $\phi_{0}$ is
the phase difference at zero magnetic field, is shown in Fig.~\ref{fig:statics} (right) as a function of the applied magnetic field. 
While in ordinary nonpolar NLCs a finite threshold field needs to be exceeded to observe a response to the external field, in the ferromagnetic case the response is thresholdless. 

\begin{figure}[htb]
\includegraphics[width=2.7in]{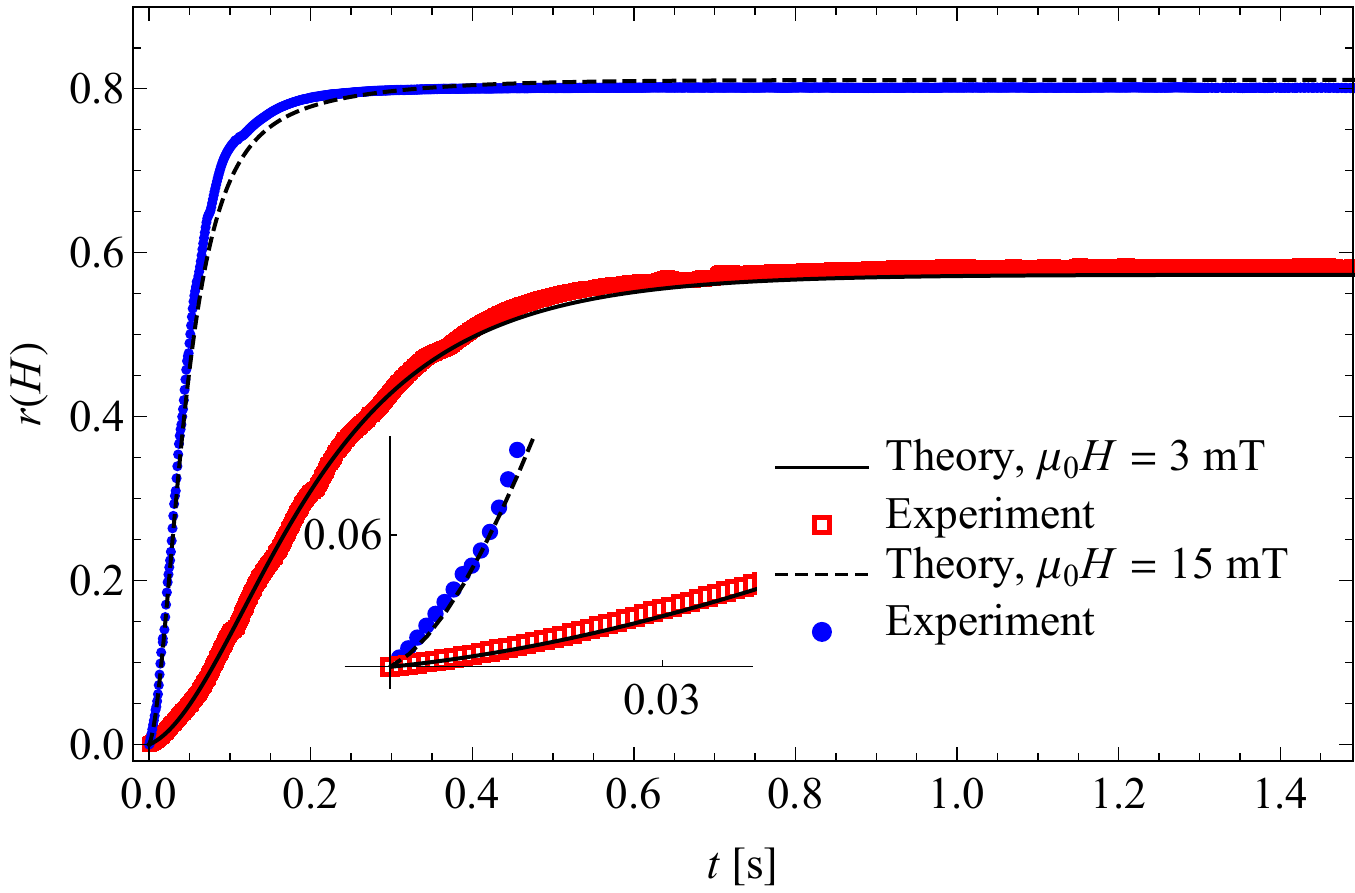} 
\includegraphics[width=2.7in]{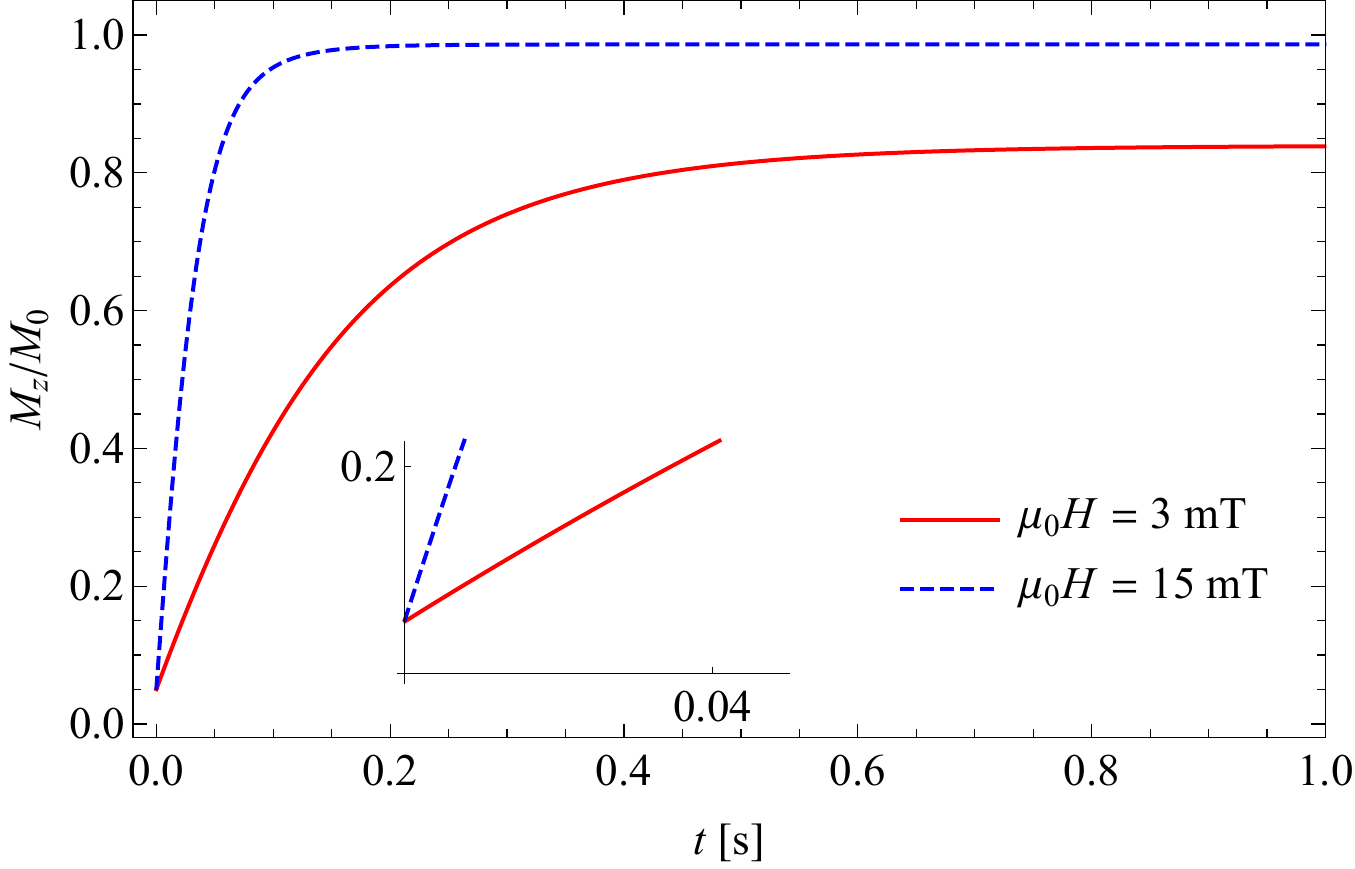} 
\caption{ (Color online) 
Top: time evolution of the normalized phase difference, $r(H)$, fitted by the dynamic model Eqs.~(\ref{f})-(\ref{bDij}).
The linear-quadratic onset of $r(H)$ is in accord with the analytic result
given in Eq.~(\ref{short}). 
Bottom: the corresponding theoretical time evolution of $M_{z}/M_{0}$, initially growing linearly as expected analytically. 
%The theoretical time dependence of the normalized $z$ component of the magnetization, $M_{z}/M_{0}$, for $\varphi_{s}=0.05$ and $\mu_{0}H=3$\,mT (solid), $\mu_{0}H=6$\,mT (dashed) and $\mu_{0}H=9$\,mT (dotted). 
}
\label{fig:dynamics} 
\end{figure}

\begin{figure}[htb]
\includegraphics[width=2.7in]{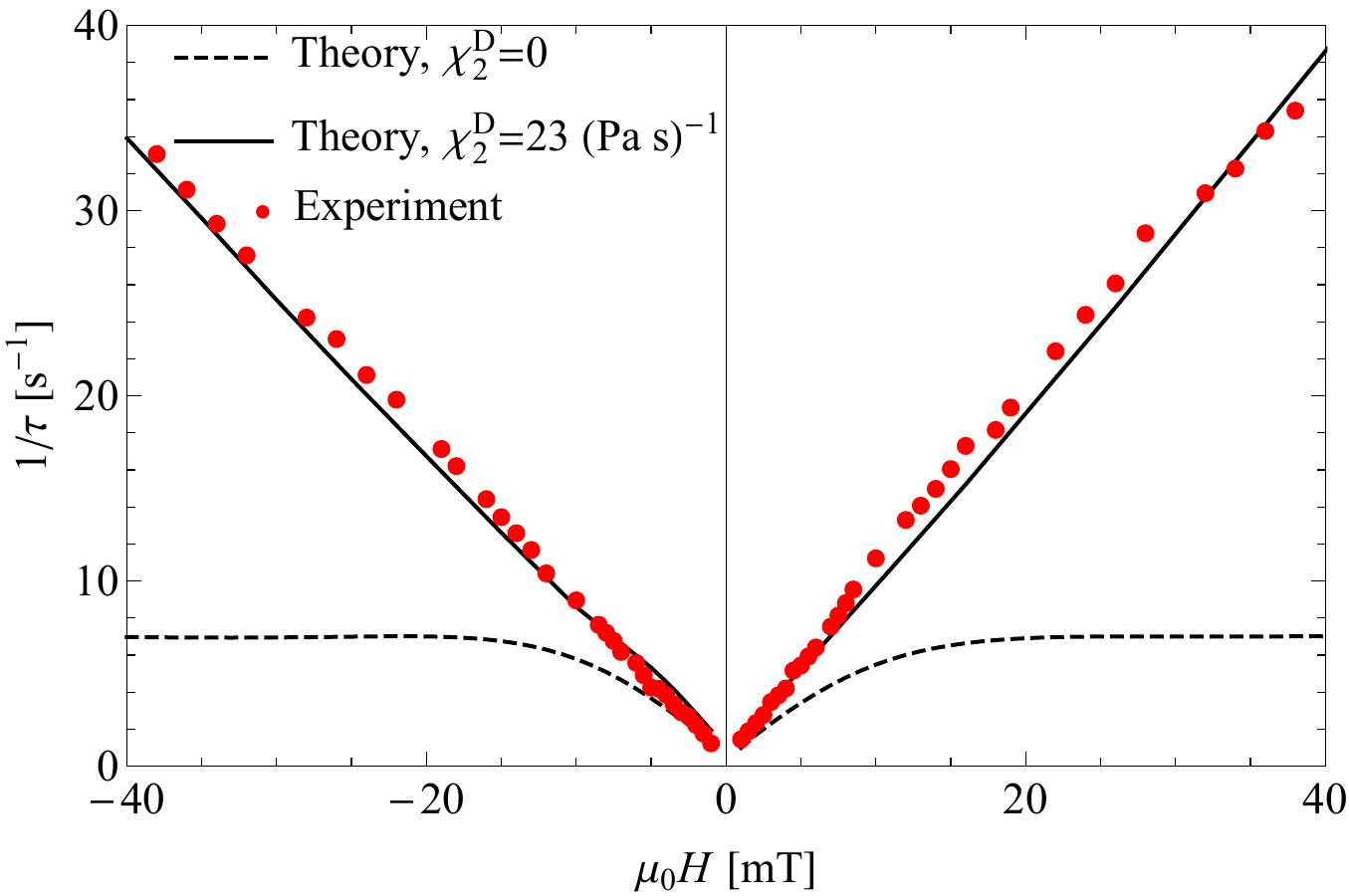} 
\caption{ 
(Color online) The inverse of the switching time, $1/\tau(H)$,
as a function of the magnetic field $\mu_{0}H$, extracted from the
experimental data and the theoretical results using the fitting function Eq.~(\ref{sigmoidal}). 
Without the dynamic cross-coupling, $1/\tau(H)$ saturates already at low fields
(dashed). 
%as the transient angle between ${\bf M}$ and ${\bf n}$ gets larger.
}
\label{fig:tauinverse} 
\end{figure}

The static response was quantitatively studied in Ref.~\cite{alenkasoftmatter}.
Here we focus on the dynamics of the response. 
Fig.~\ref{fig:dynamics} (top) shows two examples of the measured time dependence of the normalized phase difference $r(H)$. 
%at rather distinct values of the magnetic field, $\mu_{0}H=3$\,mT and $\mu_{0}H=15$\,mT.
% this statement is repeated later - there it has greater value
%
The time dependences of $r(H)$ acquired systematically for several field strengths were fitted by a squared sigmoidal function 
\begin{equation}
	f(t)=C'\left[1-\frac{1+C}{1+C\exp(-2t/\tau)}\right]^{2}\label{sigmoidal}
\end{equation}
to obtain the characteristic switching time $\tau(H)$.
Remarkably, its inverse shows a linear dependence on $H$, Fig.~\ref{fig:tauinverse}.  
Considering only a static (energetic) coupling between ${\bf M}$ and ${\bf n}$ one would expect that $1/\tau(H)$ saturates already at low fields as the transient angle between ${\bf M}$ and ${\bf n}$ gets larger.

In a minimal theoretical
model we include the magnetization field ${\bf M}({\bf r})$ and the
director field ${\bf n}({\bf r})$ and focus on the essential ingredients
of their dynamics necessary to capture the experimental results. For
a complete set of macroscopic dynamic equations for ferronematics
we refer to Ref.~\cite{jarkova} and for ferromagnetic NLCs to
Ref.~\cite{tilenlong}.

The statics is described by a free energy density $f({\bf M},{\bf n},\nabla{\bf n})$,
\begin{equation}
f=-\mu_{0}{\bf M}\cdot{\bf H}-{\textstyle \frac{1}{2}}A_{1}({\bf M}\cdot{\bf n})^{2}+{\textstyle \frac{1}{2}}A_{2}\left(|{\bf M}|-M_{0}\right)^{2}+f^{F},%
%f^{F}({\bfn},\nabla{\bfn}),
\label{f}
\end{equation}
where $\mu_{0}$ is the magnetic constant, ${\bf H}=H\hat{{\bf e}}_{z}$
is the homogeneous magnetic field fixed externally (since $H\gg M_0$),
$A_{1,2}>0$ will be assumed constant
and the Frank elastic energy of director distortions is \cite{degennesbook}
\begin{eqnarray}
f^{F} & = & {\textstyle \frac{1}{2}}K_{1}(\nabla\cdot{\bf n})^{2}+{\textstyle \frac{1}{2}}K_{2}\left[{\bf n}\cdot(\nabla\times{\bf n})\right]^{2}\nonumber \\
 & + & {\textstyle \frac{1}{2}}K_{3}\left[{\bf n}\times(\nabla\times{\bf n})\right]^{2},\label{f^F}
\end{eqnarray}
with positive elastic constants for splay ($K_{1}$), twist ($K_{2}$),
and bend ($K_{3}$). 
% Alternatively, it can be written as%$f^F = {\textstyle{1\over 2}K_{ijkl}}(\partial_j n_i)(\partial_l n_k)$, where%\begin{equation}%        K_{ijkl} = K_1 \delta_{ij}\delta_{kl} + K_2 n_p n_q \epsilon_{pij}\epsilon_{qkl} + K_3 n_j n_l \delta_{ik}.     %        \label{K}%\end{equation}
To a good approximation, one can assume that $|{\bf M}|=M_{0}$. We will
however allow for small variations of $|{\bf M}|$ (large $A_{2}$),
which is physically sound and technically convenient.

\begin{comment}
Generally, one is also interested in the coupling of gradients of
${\bf n}$ and ${\bf M}$ and should therefore in Eq.~(\ref{f})
augment the director elastic energy Eq.~(\ref{f^F}) to the complete
elastic energy functional 
\begin{eqnarray}
f^{el} & = & {\textstyle \frac{1}{2}}K_{ijkl}(\partial_{j}n_{i})(\partial_{l}n_{k})+C_{ijkl}(\partial_{j}M_{i})(\partial_{l}n_{k})\label{KC}\\
 & + & {\textstyle \frac{1}{2}}D_{ijkl}(\partial_{j}M_{i})(\partial_{l}M_{k}),\label{gradMgradM}
\end{eqnarray}
where the contributions Eq.~(\ref{gradMgradM}) must be included
as well, to ensure positivity of $f^{el}$ and thus the stability
of the homogeneous state. For fixed $|{\bf M}|$, $D_{ijkl}$ is of
the form Eq.~(\ref{K}), with the spin wave stiffness constants $D_{1,2,3}$
that are at most of the order of $0.01K_{1,2,3}/M_{0}^{2}$ \cite{alenka_softmatter},
and will be therefore omitted. The elastic coupling tensor $C_{ijkl}$
is odd both in ${\bf n}$ and in ${\bf M}$. Assuming the material
tensors in Eq.~(\ref{KC})-(\ref{gradMgradM}) are quasi isotropic,
the positivity of $f^{el}$ yields the estimate $|C|<\sqrt{KD}\sim0.1K/M_{0}$
of the restriction on the typical elements $K$, $C$, $D$ of the
tensors $K_{ijkl}$, $C_{ijkl}$, $D_{ijkl}$, respectively. The elastic
coupling contribution ($C_{ijkl}$) is thus still an order of magnitude
smaller than the director distortion energy Eq.~(\ref{f^F}) and
will be omitted as well. 
\end{comment}

At the cell plates, the director is anchored with a finite surface
anchoring energy \cite{rapini},
$
f^{S}=-{\textstyle \frac{1}{2}}W({\bf n}_{S}\cdot{\bf n})^{2}
%\label{f^S}
$,
where $W$ is the anchoring strength and ${\bf n}_{S}=\hat{{\bf e}}_{z}\sin\varphi_{s}+\hat{{\bf e}}_{x}\cos\varphi_{s}$
is the preferred direction specified by the director pretilt angle
$\varphi_{s}$. %The saddle-splay elastic energy \cite{degennes_book}%${\textstyle{1\over 2}} K_{24} {\bnu}\cdot[({\bf n}\cdot\nabla){\bf n}-(\nabla\cdot{\bf n}){\bf n}]$ %is zero in the considered geometry.

The total free energy is $F=\int\!\!f\,{\rm d}V+\int\!\!f^{S}\,{\rm d}S$
and the equilibrium condition requires $\delta F=0$.

The dynamics is governed by the balance equations \cite{pleinerbrandchapter,jarkova}
\begin{eqnarray}
\dot{M}_{i}+X_{i}^{R}+X_{i}^{D} & = & 0,\label{Mdot}\\
\dot{n}_{i}+Y_{i}^{R}+Y_{i}^{D} & = & 0,\label{ndot}
\end{eqnarray}
where the quasi-currents have been split into reversible ($X_{i}^{R}$,
$Y_{i}^{R}$) and irreversible, dissipative ($X_{i}^{D}$, $Y_{i}^{D}$)
parts. The reversible (dissipative) parts have the same (opposite)
behavior under time reversal as the time derivatives of the corresponding
variables, i.e., Eqs.~(\ref{Mdot})-(\ref{ndot}) are invariant under
time reversal if and only if dissipative quasi-currents are zero.

The quasi-currents are expressed as linear combinations of conjugate
quantities (thermodynamic forces), which in our case are the molecular fields
\begin{eqnarray}
h_{i}^{M} & \equiv & \frac{\delta f}{\delta M_{i}}=\frac{\partial f}{\partial M_{i}},\label{h^M}\\
h_{i}^{n} & \equiv & \delta_{ik}^{\perp}\frac{\delta f}{\delta n_{k}}=\delta_{ik}^{\perp}\left(\frac{\partial f}{\partial n_{k}}-\partial_{j}\Phi_{kj}\right),\label{h^n}
\end{eqnarray}
where $\Phi_{kj}={\partial f/\partial(\partial_{j}n_{k})}$ and $\delta_{ik}^{\perp}=\delta_{ik}-n_{i}n_{k}$
projects onto the plane perpendicular to the director owing to the
constraint ${\bf n}^{2}=1$.
These molecular fields can be viewed as exerting torques on {\bf M} and {\bf n}. In equilibrium they are zero, yielding the static solutions for $\bf M$ and $\bf n$, Fig.~\ref{fig:statics}.
When they are nonzero, they generate quasi-currents, which drive the dynamics through Eqs.~(\ref{Mdot})-(\ref{ndot}). 
If there is no dynamic cross-coupling, ${\bf h}^M$ drives the dynamics of $\bf M$ and ${\bf h}^n$ drives the dynamics of {\bf n}. 
In Fig.~\ref{fig:tauinverse}, $1/\tau(H)$ for this case is shown dashed. The clear deviation from the experiments indicates the importance of the dynamic cross-coupling.

We will focus on the dissipative quasi-currents as
they have a direct relevance for the explanation of the experimental
results discussed. 
%\ds{\DS{if anything, this could go out}Reversible quasi-currents \cite{jarkova} on the
%other hand should give rise to transient excursions of ${\bf M}$
%and ${\bf n}$ out of the $xz$ switching plane, and will be addressed
%in a subsequent study \cite{tilenlong}.} 
The dissipative quasi-currents
read \cite{jarkova} 
\begin{eqnarray}
X_{i}^{D} & = & b_{ij}^{D}h_{j}^{M}+\chi_{ji}^{D}h_{j}^{n},\\
Y_{i}^{D} & = & \frac{1}{\gamma_{1}}h_{i}^{n}+\chi_{ij}^{D}h_{j}^{M},\label{Y}
\end{eqnarray}
where 
\begin{eqnarray}
\chi_{ij}^{D} & = & \chi_{1}^{D}\delta_{ik}^{\perp}M_{k}n_{j}+\chi_{2}^{D}\delta_{ij}^{\perp}M_{k}n_{k},\\
b_{ij}^{D} & = & b_{\parallel}^{D}n_{i}n_{j}+b_{\perp}^{D}\delta_{ij}^{\perp}
\label{bDij}
\end{eqnarray}
%\DS{Alenka, $b^D_{ij} = {\textstyle{1\over 3}}(b^D_\parallel+2b^D_\perp)\delta_{ij}+(b^D_\parallel-b^D_\perp)(n_i n_j-{\textstyle{1\over 3}}\delta_{ij})$ is the only alternative that can be considered better in a sense (i.e. isotropic + deviatoric), but surely looks messier at first sight, so we keep the original}
and we will everywhere disregard the biaxiality of the material tensors
that takes place when ${\bf n}\nparallel{\bf M}$. We speculate that
a possible origin of this dissipative dynamic coupling between ${\bf M}$
and ${\bf n}$ is a microscopic fluid flow localized in the vicinity
of the rotating magnetic platelets.

The system Eqs.~(\ref{Mdot})-(\ref{ndot}) is discretized in the $z$ direction and solved numerically. 
%In the discrete version, the two surface points are best treated as to satisfy the same dynamic equations Eqs.~(\ref{Mdot})-(\ref{ndot}) as the internal points, with the addition of the surface anchoring energy Eq.~(\ref{f^S}) expressed as a volume density, and the divergence part of the force Eq.~(\ref{h^n}) replaced by its surface flux (the volume density thereof, again): 
%\begin{equation}
%	h_{i}^{n\,{\rm surf.}}=\delta_{ik}^{\perp}\left[\frac{\partial f}{\partial n_{k}}+\frac{1}{\Delta z}\left(\nu_{j}\Phi_{kj}+\frac{\partial f^{S}}{\partial n_{k}}\right)\right],
%\end{equation}
%where $\Delta z$ is the step size and $\bnu$ are the surface normals pointing outwards from the cell.
%
By fitting the static data to the model, Fig.~\ref{fig:statics} (right), we extract the values for the anchoring strength $W\sim2.3\times10^{-6}$\,J/m$^{2}$, 
%and $W\sim4.0\times10^{-6}$\,J/m$^{2}$, 
the pretilt angle $\varphi_{s}\sim0.05$ 
%and $\varphi_{s}\sim-0.03$ 
and the static magnetic coupling coefficient
$A_{1}\sim130\mu_{0}$.
The agreement between the static experimental data and the model underscores that we have solid ground for the analysis
of the dynamics.

In Fig.~\ref{fig:dynamics} (top) the measured time dependence of the normalized
phase difference $r(H)$ is compared \cite{alenkasoftmatter}
 to the model for two rather distinct values of the magnetic field.
%, $\mu_{0}H=3$\,mT and $\mu_{0}H=15$\,mT. 
The fits are performed by varying the values of
the dynamic parameters subject to stability restrictions, while keeping
the values of $W$, $\varphi_{s}$, and $A_{1}$ fixed as
determined from the statics. The model captures the dynamics very
well for all times from the onset to the saturation. The extracted
values of the dynamic parameters are $\gamma_{1}\sim0.03$\,Pa\,s,
$b_{\perp}^{D}\sim7.8\times10^{4}$\,Am/Vs$^2$ and $\chi_{2}^{D}\sim23$\,(Pa\,s)$^{-1}$,
which safely meets the positivity condition of the entropy production
$|\chi_{2}^{D}|<\sqrt{b_{\perp}^{D}/(\gamma_{1}M_{0}^{2})}\sim32$\,\,(Pa\,s)$^{-1}$.
The remaining two dynamic parameters do not affect the dynamics significantly
and are set to 
%$b_{\parallel}=1\times10^{5}$\,Am/Vs$^2$
$b^D_{\parallel}=b^D_\perp$
and 
$\chi_{1}^{D}=0$.
%$\chi_{1}^{D}=\chi_{2}^{D}$.

Fig.~\ref{fig:dynamics} (bottom) shows the corresponding theoretical time dependence of the normalized $z$
component of the magnetization, which is not measured due to insufficient 
time resolution of the vibrating sample magnetometer 
\cite{alenkasoftmatter} (LakeShore 7400 Series VSM, several
seconds are required for ambient magnetic noise averaging).
%For small times the magnetization grows linearly, a result that is also obtained analytically. 
%As one expects intuitively the rise time for the magnetization is reduced as the applied magnetic field is increased. 

Initially, ${\bf n}$ is homogeneous and aligned with ${\bf M}$, 
such that ${\bf h}^n$ is zero and the director dynamics in Eq.~(\ref{Y}) is due to ${\bf h}^M$ alone.
For small times when ${\bf M}$ and ${\bf n}$ are only slightly pretilted from the $x$ direction,
it thus follows from Eqs.~(\ref{ndot}) and (\ref{Y}) that
\begin{equation}
\label{nzinit}
	n_{z}(t)\approx\varphi_{s}+\chi_{2}^{D}M_{0}\mu_{0}H\,t
\end{equation}
and hence {\cite{alenkasoftmatter}}
\begin{eqnarray}
\label{short}
r(H)\hspace{-2mm} & = & \hspace{-2mm}\frac{n_{0e}(n_{0e}+n_{0})}{2n_{0}^{2}}[(\chi_{2}^{D}M_{0}\mu_{0}H)^{2}t^{2}+2\varphi_{s}\chi_{2}^{D}M_{0}\mu_{0}Ht]\nonumber \\
 & \equiv & k^{2}t^{2}+pt,
\end{eqnarray}
which is also revealed by Fig.~\ref{fig:dynamics} (top, inset);
$n_{0}$ and $n_{0e}$ are the ordinary and the extraordinary refractive
indices. 
In principle, $k^2$ contains a small static coupling ($A_1$) correction linear in the pretilt, which is however within the error margin and is neglected. 
The initial dynamics of the director and the behavior of
$r(H)$ are thus governed by the dissipative dynamic cross-coupling
between director and magnetization, Eq.~(\ref{Y}), described by
the parameter $\chi_{2}^{D}$. 

%Compare this to the situation where the cross-coupling were absent, resulting in \cite{tilenlong} 
%\begin{equation}
%	n_{z}(t)\approx\varphi_{s}+\frac{A_{1}M_{0}b_{\perp}^{D}\mu_{0}H}{2\gamma_{1}}\,t^{2}
%\end{equation}
%and the phase difference starting as $\propto t^{4}$ in case of zero pretilt.

\begin{figure}[htb]
\includegraphics[width=2.7in]{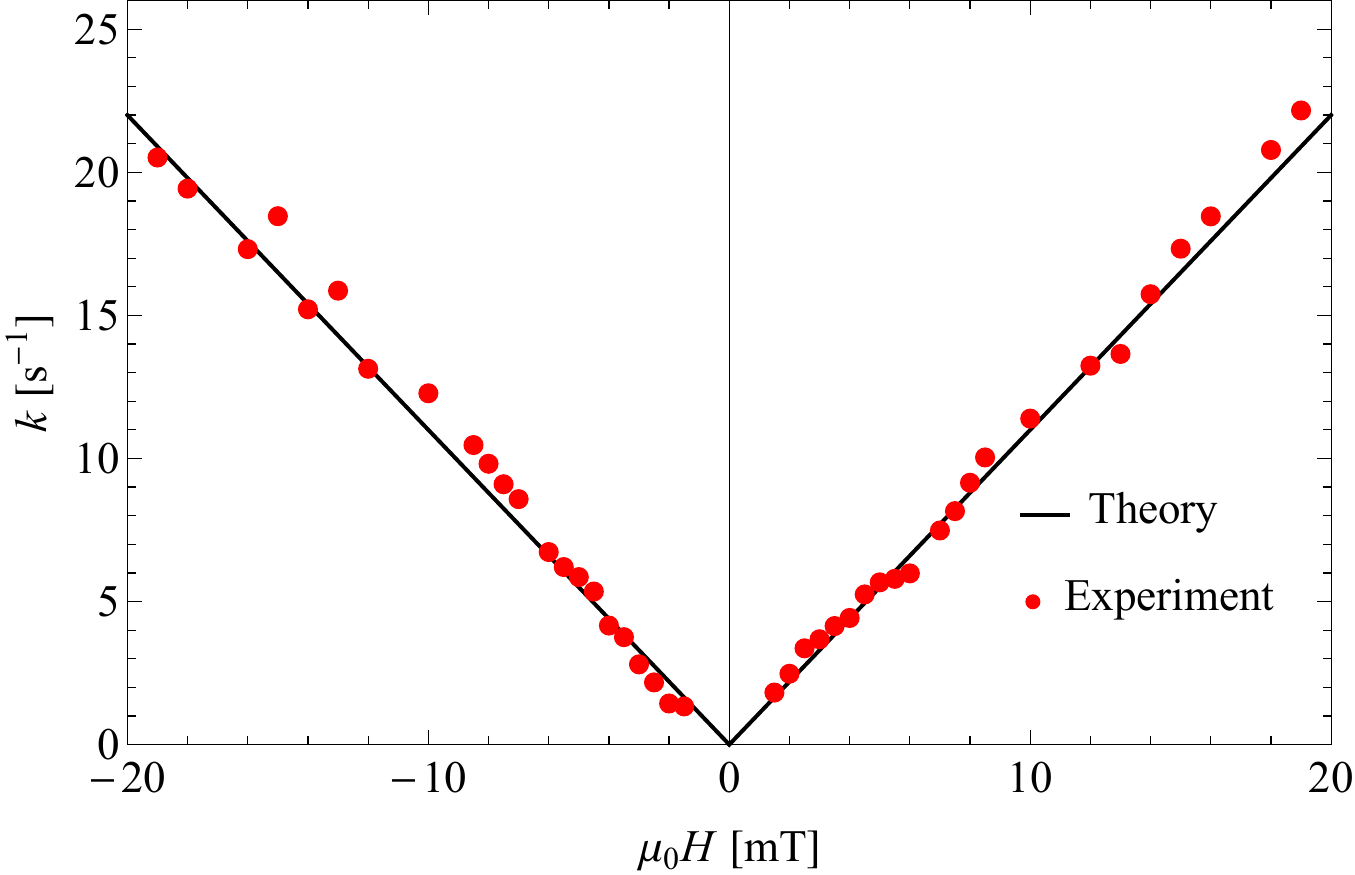} \caption{ (Color online) The 
coefficients $k$ and $p$ (inset) of Eq.~(\ref{short}) as functions
of $\mu_{0}H$, extracted from the initial stage $r(H)$ 
measurements by the straight line fits.}
\label{fig:initial} 
\end{figure}

Fitting Eq.~(\ref{short}) to the initial time evolution of measured
normalized phase differences for several values of the magnetic field
$\mu_{0}H$, we determine the parameters $k$ and $p$, Fig.~\ref{fig:initial},
and extract therefrom values of the dissipative 
magnetization-director coupling parameter
$\chi_{2}^{D}\sim (21\pm2)$\,(Pa\,s)$^{-1}$ and 
the pretilt $\varphi_{s}\sim0.05\pm0.03$.

The best match of $1/\tau(H)$, Fig.~\ref{fig:tauinverse}, extracted from the
experimental data and the model via Eq.~(\ref{sigmoidal}), 
allows for a robust evaluation of the
dissipative magnetization-director coupling parameter:
$\chi_{2}^{D}=(23\pm2)$\,(Pa\,s)$^{-1}$.
The theoretical results confirm that the linear shape of 
$1/\tau(H)$ is due precisely to this dissipative cross-coupling
and would not take place if only the static coupling (the $A_{1}$
term in Eq.~(\ref{f})) were at work, as demonstrated in Fig.~\ref{fig:tauinverse} (dashed curve).

The coupling of $\bf M$ and ${\bf n}$ to flow was not taken into account.
As flow is generated by gradients (i.e., divergence of the stress tensor), 
starting with a homogeneous configuration it is absent initially.
To lowest order, Eq.~(\ref{nzinit}) is thus unaffected by the flow coupling, irrespective of its details.
Moreover, in ordinary NLCs the small backflow effect 
%(coupling of $\bf n$ and flow) 
makes the response a little faster \cite{danielzumer}. In a ferromagnetic NLC, 
additional couplings to the velocity field are possible.
% the effects of which have yet to be studied. 
Nevertheless, the match of $\chi_{2}^{D}$ extracted from the initial (where flow is absent) and the overall dynamics speaks for only a minor flow coupling effect.

In summary, we have presented experimental and theoretical investigations
of the magnetization and director dynamics in a ferromagnetic
liquid crystal. We have demonstrated that a dissipative cross-coupling
between the magnetization and the director, which has been determined quantitatively,
is crucial to describe the experimental results. 
Such a coupling arises for all systems with macroscopic magnetization 
and director fields and its presence dictated by symmetry has been 
pointed out before for ferronematics. Clearly its strength is expected to 
be higher in ferromagnetic systems of the type studied here.
This coupling makes the response of such materials much faster, which is important 
for potential applications
in magneto-optic devices, e.g., devices for magnetic field visualization \cite{visualization}. 
Their main advantage compared to existing techniques is that both the magnitude and the direction of the field can be simultaneously visualized. 
Further possible applications include
remote optical sensing of magnetic fields and the use of a magnetic field to manipulate complex (patterned) structures in liquid crystals, e.g. for spatial light modulation \cite{smalyukhAPL,smalyukhNatMat}. 
An advantage of the magnetic field is that it can be applied in a non-contact way in any direction, whereas the application of the electric field is limited by the geometry of the electrodes.
The main challenge is to produce a variety of suspensions with different magnetic and viscoelastic properties, stable in a wide temperature range.

We have laid
here, in a pioneering step, the experimental and theoretical basis
of a dynamic description. 
Naturally to include the coupling to flow
is next. First experimental results in this direction have been described
in Ref.~\cite{alenkaapl}, where it was shown that viscous effects
can be tuned by an external magnetic field of about $10^{-2}$\,T
by more than a factor of two, indicating a potential for applications in the field of smart fluids.

%Thus low-magnetic-field effects in ferromagnetic NLCs have potential for applications as magneto-optic devices as well as in the field of smart fluids \cite{visualization}.

As ferromagnetic NLCs have two order parameters characterized
by the magnetization and the director field, 
%they represent an example of a fluid multiferroic. As such 
they offer the possibility to chiralize
the material to obtain a ferromagnetic cholesteric NLC breaking
parity and time reversal symmetry in a fluid ground state. 
The formation of solitons in an unwound ferromagnetic cholesteric NLC has been recently realized and discussed in Refs.~\cite{smalyukhPRL,smalyukhNatMat}.
Another
promising direction to pursue will be to produce a liquid crystalline
version of uniaxial magnetic gels \cite{collin2003,bohlius2004}.
Cross-linking a ferromagnetic NLC gives rise to the possibility
to obtain a soft ferromagnetic gel opening the door to a new class
of magnetic complex fluids. This perspective looks all the more promising
since recently \cite{menzel2014,gka2016} important physical properties
of magnetic gels such as nonaffine deformations \cite{menzel2014}
and buckling of chains of magnetic particles \cite{gka2016} are characterized
well experimentally and modeled successfully.

%\vspace{-6mm}

%\DS{could we shorten the acknowledgment?}
Partial support through the Schwerpunktprogramm SPP 1681 
%'Feldge\-steuerte
%Partikel-Matrix-Wechselwirkungen: Erzeugung, skalen�bergreifende Modellierung
%und Anwendung magnetischer Hybridmaterialien' 
of the Deutsche Forschungsgemeinschaft
is gratefully acknowledged by H.R.B.,H.P., T.P. and D.S., 
as well as the support of the Slovenian
Research Agency, Grants N1-0019, J1-7435 (D.S.), P1-0192 (A.M. and N.O.) and P2-0089 (D.L.).

% \vspace{-0.5cm}

\end{document}